\documentstyle[prl,twocolumn,aps,graphicx,epsfig,amssymb]{revtex}
\begin{document}
\draft
%\begin{frontmatter}
\title{Hydration of Methanol in Water\\ A DFT-based Molecular Dynamics Study}
\author{Titus S. van Erp and Evert Jan Meijer\\
Department of Chemical Engineering,
Universiteit van Amsterdam,
Nieuwe Achtergracht 166, NL-1018 WV AMSTERDAM, The Netherlands}

\maketitle
\vspace*{-10mm}

\begin{abstract}
\noindent
  We studied the hydration of a single methanol molecule in aqueous
  solution by first-principle DFT-based molecular dynamics simulation.
  The calculations show that the local structural and short-time
  dynamical properties of the water molecules remain almost unchanged
  by the presence of the methanol, confirming the observation from
  recent experimental structural data for dilute solutions. We also
  see, in accordance with this experimental work, a distinct shell of
  water molecules that consists of about 15 molecules. We found no
  evidence for a strong tangential ordering of the water molecules in
  the first hydration shell.
\end{abstract}
%%%%%%%  PREPRINT HEADER %%%%%%%%%%%%%%%%%%%%%%%%%%%%%%%%%%%%%%%%%%%%%%%%%%%%%%%
%\setlength{\unitlength}{1mm}
%\begin{picture}(20,00)(5,-65)
%  \put(0,20){\framebox(33,12)[c]{\Large\bf\sf PREPRINT}}
%  \put(37,29){\makebox(0,0)[l]{\small\sf Accepted for publication in
%  {\it Chem.\ Phys.\ Lett.}}}
%\end{picture}
%%%%%%%%%%%%%%%%%%%%%%% END OF PREPRINT HEADER %%%%%%%%%%%%%%%%%%%%%%%%%%%%
%\begin{keyword}
%\end{keyword}
%\end{frontmatter}
%\begin{multicols}{2}

\vspace*{2mm}
\centerline{\bf INTRODUCTION}

The solvation of alcohols in water has been studied
extensively.\cite{Franks} It is of fundamental interest in physics,
chemistry and biology, but also of importance in technical
applications. The characteristic hydroxyl group allows alcohols to form
hydrogen bonds and is responsible for the good solubility of the smaller
alcohols. In contrast, the alkyl group is hydrophobic and does not
participate in the hydrogen bonding network of water. The presence of
both hydrophobic and hydrophilic groups make the microscopic picture
of solvation of alcohol in water a non-trivial and therefore
interesting matter.

Understanding the solvation of methanol in water is a prerequisite for
the study of chemistry of alcohols in aqueous solution.  Important
examples of such reactions are the conversion of ethanol into
acetaldehyde in biological systems or the industrial ethanol
production by acid-catalysed hydration of ethylene.  An accurate
microscopic understanding of the mechanism and kinetics of such
reactions is of fundamental interest. However, presently, this picture
is still far from complete.  Density Functional Theory (DFT) based
Molecular Dynamics simulation has proved to be a promising tool
provide such an insight. An accurate calculation of the chemical
bonding is incorporated via a DFT-based electronic structure
calculations. The effect of temperature and solvent on the reactive
events is implicitly accounted for via the Molecular Dynamics
technique. The implementation of DFT-based MD as proposed by Car and
Parrinello\cite{CaPa85} has proven to be extremely efficient. It has
successfully been applied to study of a large variety of
condensed-phase systems at finite temperature. Applications to
chemical reactions include the cat-ionic polymerization of
1,2,5-trioxane\cite{CuAn94}, or the acid-catalysed hydration of
formaldehyde\cite{MeSp98}.

As a first step towards the study of chemical reactions involving
alcohols we present in this paper a Car-Parrinello Molecular Dynamics
(CPMD) study of the hydration of the simplest alcohol (methanol) in
aqueous solution.  Recent experimental work\cite{SoFi93} has provided
detailed structural information on the solvation shell. Various
molecular simulation studies (e.g.
Ref.~\cite{JoMa83,FeHa90,PaBa91,TaGu97,LaKu97} have addressed
structure and dynamics of both the solute and the solvent.  This
experimental and numerical work has revealed that there is a distinct
solvation shell around the methanol, and that the water structure is
little affected by the presence of a methanol molecule. In this paper
we will address these structural properties and in addition consider
the dynamics of the methanol and the water molecules in the solvation
shell.

This paper is organized as follows.  First we outline the computational
approach and its validation. Then we present the results for the
structure and dynamics of a single solvated methanol in water.
We conclude the paper with a summary and discussion.
\vspace*{2mm}

\centerline{\bf METHODS AND VALIDATION} Electronic structure
calculations are performed using the Kohn-Sham
formulation\cite{KoSh65} of DFT.\cite{HoKo64} We employed the BLYP
functional, that combines a gradient-corrected term for the
correlation energy as proposed by Lee, Yang and Parr\cite{LeYa88} with
the gradient correction for the exchange energy due to
Becke\cite{Beck88_2}.  Among the available functionals, the BLYP
functional has proven to give the best description of the structure
and dynamics of water.\cite{SpHu96,SiBe97} All
calculations\cite{CompResource} were performed using the CPMD
package.\cite{CPMD30f}

The pseudopotential method is used to restrict the number of
electronic states to those of the valence electrons. The interaction
with the core electrons is taken into account using semi-local
norm-conserving Martins-Troullier pseudopotentials.\cite{TrMa91} The
pseudopotential cutoff radius for the H was chosen 0.50 au. For O and
C the radii are taken 1.11 and 1.23 a.u. for both the l=s and l=p
term.  The Kohn-Sham states are expanded in a plane-wave basis set
matching the periodicity of the periodic box with waves up to a kinetic
energy of 70~Ry.  Test calculations showed that for this structural
and energetic properties were converged within $0.01~$\AA\ and $1
$~kJ/mol, respectively. Frequencies are converged within 1~\%, expect
for CO and OH stretch modes that are underestimated by 3~\% and 5~\%
compared to basis-set limit values.

\begin{figure}
\begin{center}
\includegraphics[angle=0,width=4cm]{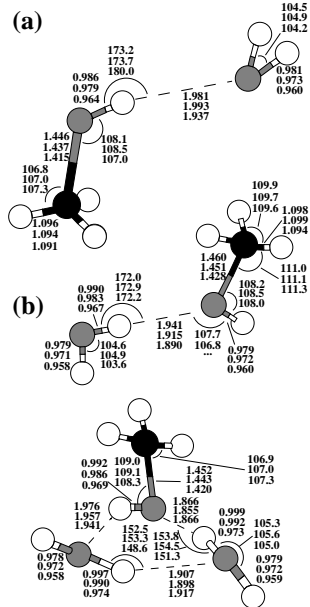}
\caption{
\label{fig:complex}
  Energy-optimized geometries of two water$/$methanol dimers and a
  trimer. Distances (\AA) and angles (degrees) are shown for three
  computational methods: CPMD-BLYP (top, present work),
  ADF-BLYP\protect\cite{ADF23} (middle, present work) and MP2
  \protect\cite{GoMo98} (bottom).}
\end{center}
\end{figure}

To validate the computational methods outlined above we performed a
series of reference calculations of relevant gas-phase compounds with
the CPMD package. Energetics and geometry were calculated for
methanol, water, two mono-hydrate configurations, and the di-hydrate
configuration shown in Fig.~\ref{fig:complex}.  These calculations
were performed using a a large periodic box of size 10x10x10~\AA$^3$.
The interactions among the periodic images were eliminated by a
screening technique similar to that of Ref.~\cite{BaLa93}. In addition
we determined for the methanol molecule both the harmonic vibrational
frequencies and the frequencies at finite temperature (T=~200~K).  The
latter includes the anharmonic contributions, and were obtained from
the spectrum of the velocity auto correlation function (VACF) of a
3~ps CPMD calculation at E=~200~K.  The calculated peak positions can be
compared with experimental spectra.  Results of the gas-phase
calculations were compared with results obtained with a
state-of-the-art atomic-orbital based DFT package (ADF\cite{ADF23}),
and with results from MP2 calculations of Ref.~\cite{GoMo98}. In the
comparison of the energies zero-point energies were not taken into
account.

\vspace*{-3mm}
\vspace*{-4mm}

Complexation energies and geometries of the methanol hydrates are
given in Tab.~\ref{tab:E_complex} and Fig.~\ref{fig:complex}.
Deviations among CPMD and ADF are within 1~kcal/mole for the energies,
smaller than $0.005$~\AA\ for the inter-molecular bonds and within
$0.03$~\AA\ for the weaker intra-molecular bonds. This indicates a
state-of-the art accuracy for electronic structure methods employed in
CPMD.  Differences among BLYP and MP2 are within acceptable limits,
with BLYP complexation energies smaller by 4~kJ/mole (dimer) and
10~kJ/mole (trimer).  These deviations are similar to the comparison
of BLYP and MP2 for the water dimer binding
energy,\cite{SpHu96,MP2limit} where BLYP is 4~kJ/mole smaller, with
the MP2 energy only $1$~kJ/mol below the experimental value. Assuming
similar differences for the complexation energies bonds in the
methanol hydrates would suggest that BLYP underestimates the
methanol-water binding energy by approximately $5$~kJ/mol.  Inter- and
intra-molecular BLYP bond lengths are up to $0.02~$ and 0.06~\AA\ 
longer compared to the MP2 results, respectively.

\begin{table}[b]
\caption{\label{tab:E_complex}
Complexation energies (kJ/mol) of methanol hydrates shown in
Fig.~\ref{fig:complex}. Numbers are
bare values without zero-point energy corrections and
entropy contributions.}
\begin{tabular}{l|ccc}
Complex &  CPMD-BLYP  & ADF-BLYP$^a$ & MP2$^b$  \\
 \hline
CH$_3$O + H$_2$O (a)  & 20.2 & 20.2  & 24.4 \\
CH$_3$O + H$_2$O (b)  & 17.1 & 17.6  & 21.0 \\
CH$_3$O + 2 H$_2$O    & 58.3 & 59.6  & 68.8 \\
\end{tabular}
{\small
$^a$ Ref.~\cite{ADF_basis}.\\
$^b$ G2(MP2) method. MP2(full)/6-311+G(d,p) optimized geometries. From Ref.~\cite{GoMo98}.
}
\end{table}

Vibrational frequencies are listed in Tab.~\ref{tab:freq}.  Again
comparison of CPMD and ADF is excellent, consistent with the results
for the energetics and geometries. Comparing the calculated
finite-temperature frequencies against the experimental values shows
that BLYP tends to underestimate the frequencies of almost all modes
by $\approx$~10~\%. This trend is a known feature of BLYP. For example
similar deviations are observed for BLYP calculation of
water.\cite{SpHu96}

Overall we conclude that the reference calculations of gas-phase
provides confidence that DFT-BLYP performs with a sufficient accuracy
for a quantitative study of methanol hydration.\\

\begin{table}
\caption{
Harmonic and T=200~K vibrational frequencies of
gas-phase methanol molecule.}
\begin{tabular} {l|cc|cc}
 & \multicolumn{2}{c|}{Harmonic} &
   \multicolumn{2}{c}{Anharmonic}\\
 & \multicolumn{2}{c|}{$\nu$ (cm$^{-1}$)} &
   \multicolumn{2}{c}{$\nu$ (cm$^{-1}$)}\\
mode &  {\footnotesize CPMD-BLYP}  &
 {\footnotesize ADF-BLYP$^a$} &{\footnotesize CPMD-BLYP} &
 {\footnotesize Exp.$^b$}\\
& & & {\footnotesize (T=200~K)} & \\
 \hline
 \vspace{-3mm} &&&& \\
$\tau$(OH)       &  280 &  380 &   280      &   270 \\
$\nu$(CO)        &  940 &  950 &   880      &  1034 \\
$r$(CH$_3$)      & 1040 & 1050 &   980      &  1075 \\
$r$(CH$_3$)      & 1130 & 1130 &  1070      &  1145 \\
$\delta$(OH)     & 1330 & 1340 &  1270      &  1340 \\
$\delta$(CH$_3$) & 1430 & 1430 &  1320-1430$^c$ &  1454 \\
$\delta$(CH$_3$) & 1460 & 1460 &  1320-1430$^c$ &  1465 \\
$\delta$(CH$_3$) & 1470 & 1470 &  1320-1430$^c$ &  1480 \\
$\nu$(CH$_3$)    & 2940 & 2910 &  2640      &  2844 \\
$\nu$(CH$_3$)    & 2990 & 2950 &  2740      &  2970 \\
$\nu$(CH$_3$)    & 3060 & 3020 &  2830      &  2999 \\
$\nu$(OH)        & 3550 & 3590 &  3310      &  3682 
\end{tabular}
{\small
$^a$ Ref.~\cite{ADF_basis}.\\
$^b$ Ref.~\cite{BeGa56}.\\
$^c$ Modes not separated. Broad peak with width listed.
}
\label{tab:freq}
\end{table}

\centerline{\bf SOLVATION}

We performed Car-Parrinello Molecular Dynamics simulations of the
solvation of a single methanol molecule. We considered two systems:
one with 31 water molecules and the other with 63 water molecules,
yielding methanol-water solutions with mole ratios of 1:31 and 1:63.
In the following they are referred to as the small and large system,
respectively.  For reference we also performed a simulation of a pure
water sample of 32 molecules. The molecules are placed in a periodic
cubic box with edges of 9.98~\AA\ (small solvated methanol system),
12.50~\AA\ (large solvated methanol system), and $9.86~\AA$ (pure
water) corresponding to the experimental densities at ambient
conditions.  The temperature of the ions is fixed at 300~K using a
Nos\'e-Hoover thermostat \cite{Nose84_1,Nose84_2,Hoov85}.  The
fictitious mass associated with the plane-wave coefficients is chosen at
900 a.u., which allowed for a time step in the numerical integration of
the equations-of-motion of 0.145~fs.  The two systems were
equilibrated for 1 ps from an initial configuration obtained by a
force-field simulation. Subsequently we gathered statistical averages
from a 10~ps trajectory of the 31+1 molecule system, from a 7~ps
trajectory of the 63+1 molecule system, and from a 10~ps trajectory
of the pure water system.
\vspace*{2mm}

\centerline{\bf Structure}

\begin{figure}[h!]
\includegraphics[angle=0,width=8cm]{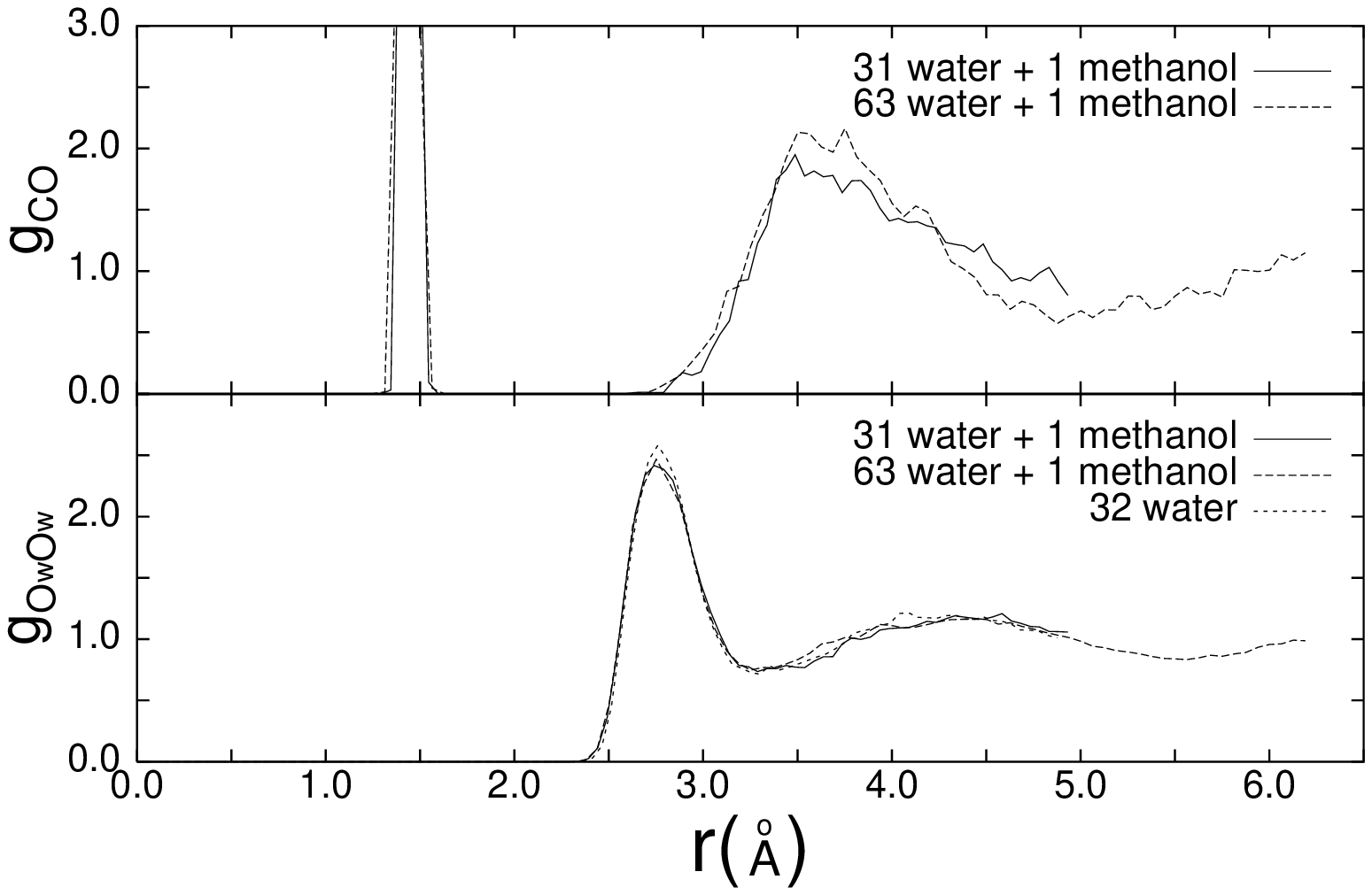}
\vspace*{5mm}
\caption{
\label{fig:RDF}
Calculated carbon-oxygen (top) and water oxygen-oxygen (bottom) radial distribution functions for
  the small (solid line) and large (dashed line) methanol system.}
\end{figure}

In Fig.~\ref{fig:RDF} we have plotted the radial distribution
functions (RDF) of the water oxygen atoms.  The minor variations among
the RDF's of the small methanol system, the large methanol system, and
the pure water system is an indication that the local water structure, as
measured by this RDF, is at only marginally changed by the
solvation of a methanol molecule.  Note, in this respect, that for the
32 molecule the first solvation shell constitutes a significant
fraction of the total number of water molecules (see below).

Fig.~\ref{fig:RDF} also shows the RDF of the methanol carbon and water
oxygens for the small and large methanol system.  A pronounced first
peak clearly indicates the existence of shell of water molecules at a
distance of $\approx$3.7~\AA.  Comparing the RDF's of the small and
large system shows a noticeable difference.  This should be
attributed to the limited size of the small system. It suggests that a
proper description of the solvation structure of a single methanol in a
cubic periodic simulation box requires at least 50 water molecules.
Integrating the RDF for the large system up to the minimum at
$r=5.0$~\AA\ yields $16$ water molecules in the first solvation shell.
The definite solvation shell observed in our simulations is consistent
with the neutron diffraction data of Soper and Finney\cite{SoFi93} who
studied a 1:9 molar methanol-water system.  Differences in molarity
limits a quantitative comparison of the carbon-oxygen RDF, but a
qualitative comparison learns that peak positions match with the peak
values slightly more pronounced in the simulation results.

To analyze the orientational ordering of the water molecules around
the methanol we computed the distribution function of the angle
between the C-O$_{\rm H}{_2}_{\rm O}$ bond vector and the normal to
the plane of the water molecules in the first solvation shell. The
results show that angle distribution is relatively uniform with a
small tendency towards the tangential orientation, a feature occurs
for all solvation shell radii in the range of 3.7-5.0~\AA. Over the
range of 0$^o$-90$^o$ the distribution gradually decays, with the
value at the tangential orientation (0$^o$) about a factor of 2 larger
than at the perpendicular orientation (90$^o$).  Qualitatively, this
seems consistent with data for the orientational distribution obtained
from neutron-diffraction data \cite{SoFi93}.  However from this
experimental data it is concluded that the water molecules prefer to
lie tangential and form a cage around the methanol. Our data do not
give clear evidence for a cage-like structure.  However, this might be
a different interpretation from similar data.  Note, in this respect,
also that the experimental data cannot be quantitatively compared to
our data, as different orientational distribution functions are
employed.

To analyze the hydrogen bonding we adopted the definition of
Ref.~\cite{FeHa90}: two molecules are hydrogen bonded if
simultaneously the inter-oxygen distance is less than $3.5~$\AA
and the OHO angle
is smaller than $30^o$.  From the simulation of the large system
we found that the methanol hydroxyl group donates and accepts on
average 0.9 and 1.5  hydrogen bonds, respectively. For a
water molecule these numbers are equal and measured to be 1.7 in
the simulation of the pure water sample.  These results indicate that
the methanol hydroxyl group participates strongly in the hydrogen
bonding network with the a donating behavior similar to water
hydrogen and a accepting character somewhat smaller than a water
oxygen.
\vspace*{2mm}

\centerline{\bf Dynamics}

The time scale (7-10 ps) of the present simulations allows for a
reliable analysis of dynamical properties occurring on the picosecond
time scale. 

\begin{figure} 
\includegraphics[angle=0,width=8cm]{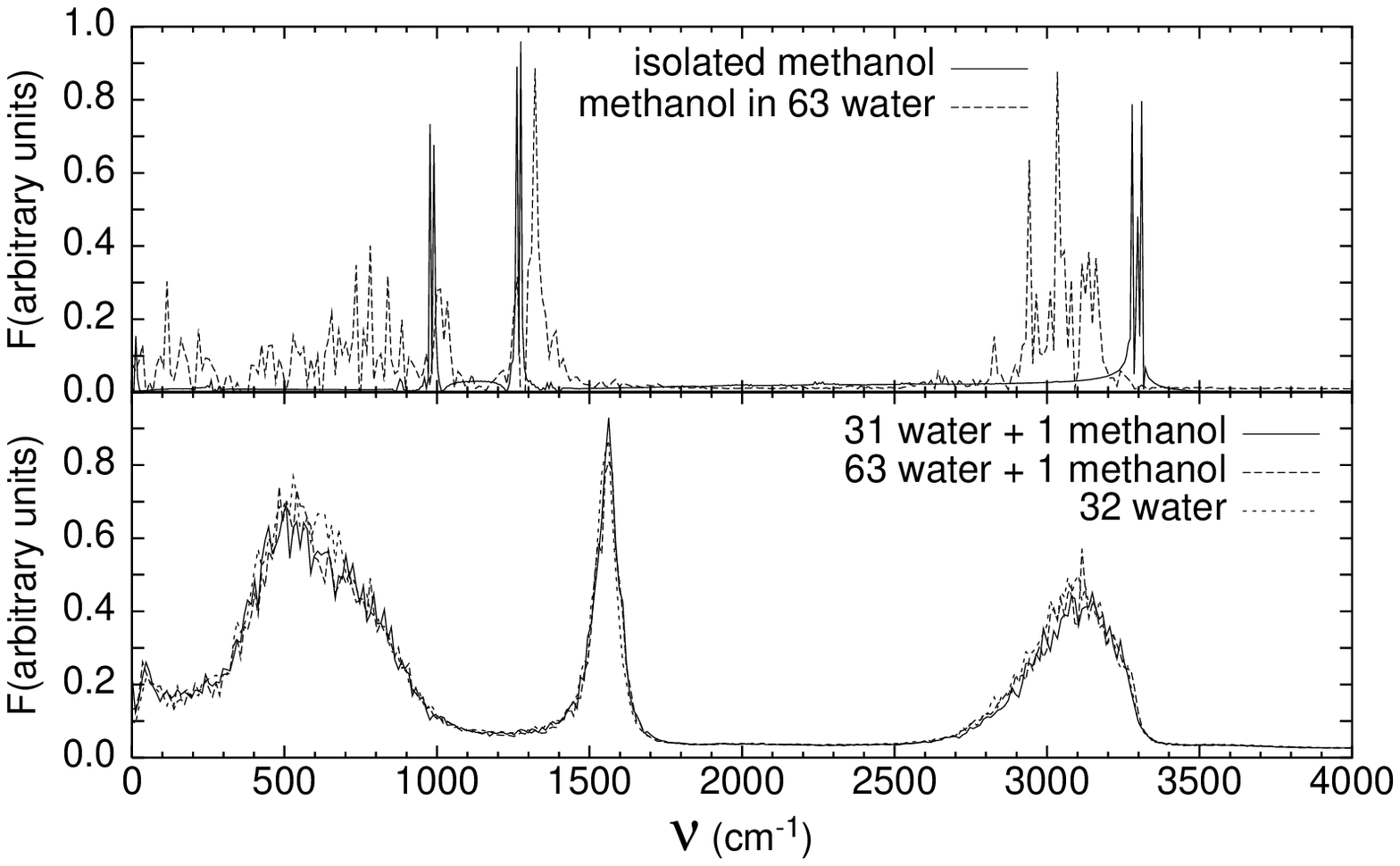}
\vspace*{3mm}
\caption{
\label{fig:VACF}
Bottom: Calculated Fourier spectrum of the velocity auto correlation
function of the water hydrogens for the small methanol system (solid
line), the large methanol system (dashed line), and the pure water
sample (dotted line).  Top: Calculated Fourier spectrum of the velocity
auto correlation function of the hydrogen atom of the methanol
hydroxyl group for the large methanol system (dashed line) and for an isolated
methanol molecule (solid line).}
\end{figure}

The velocity auto correlation function (VACF) of the hydrogen atoms
provides an important measure of hydrogen bonding.
Fig.~\ref{fig:VACF} shows the Fourier spectrum of the calculated VACF
of hydrogen atoms of the water molecules in the small and large
methanol sample.  The three distinct peaks correspond to the
vibrational (3100~cm$^{-1}$), bending (1600~cm$^{-1}$), and
librational-translational (500~cm$^{-1}$) modes of the water
molecules. The most important observation is that mutual comparison of
the two methanol samples and the comparison of these with the spectrum
of the pure water sample (also plotted) shows no significant
difference, not even for the small methanol sample where the solvation
shell constitutes half of the water molecules in the system. This
demonstrates that also the short-time dynamics of the water molecules
is hardly affected by the solvation of a methanol molecule.

An indication for the average residence time of a water molecule in the
first solvation shell is obtained by monitoring the trajectories of
the individual water molecules. We found that in the large methanol
system over 7 ps 10 water molecules left the region within $5~$\AA\
from the methanol carbon. From this we estimate the average residence time to
be of the order of a few picoseconds.

Fig.~\ref{fig:VACF} shows the Fourier spectrum of the VACF of the
hydroxyl H of methanol obtained from the trajectory of the large
system. The spectrum is of limited accuracy due to the relatively
short trajectories (7~ps).  For comparison, the calculated spectrum
for a single methanol molecule at $T=200~K$ is also plotted. In
solution the OH stretch ($\nu_{\rm OH}$) peak, with a calculated
gas-phase position of about 3300~cm$^{-1}$, has shifted by
$\approx$~200~cm$^{-1}$ to lower frequencies and has a relatively
large width. The shift and width are both typical characteristics of a
hydrogen bond and are also observed in the water spectrum
(Fig.~\ref{fig:VACF}).  In contrast to the OH stretch mode, we see
that the OH-bending  mode ($\delta_{\rm OH}$ at 1300~cm$^{-1}$) is
blue-shifted by an amount of 50-100$^{-1}$.  A comparison with
experimental frequency shifts in infrared spectra is
limited as, to our knowledge, no experimental data for dilute methanol-water
solutions are reported.  However, a comparison with measured shifts in
liquid methanol\cite{Shim72} shows similar trends for the shift of infrared
stretch (-354~cm$^{-1}$) and bend (+78~cm$^{-1}$) peaks.  The
torsional mode ($\tau_{\rm OH}$), expected to be shifted upward to
around 600~cm$^{-1}$, is not visible in our calculated spectra due to
the large statistical errors.
\vspace*{2mm}

\centerline{\bf DISCUSSION}
We have studied the solvation of a single methanol molecule in water
using DFT-based Car-Parrinello molecular dynamics simulation.
Validation of the approach showed that energetics, structural, and
dynamical properties of reference gas-phase compounds were sufficient
to expect a quantitative accuracy of calculated properties.
 
The calculated solvation structure supports the experimental
observation\cite{SoFi93} that a shell of about 15 water molecules is
formed around the methanol. Structural analysis also learns that the
hydrogen bonded network of water is only minimally distorted by the
presence of the methanol molecule. This confirms the proposition of
Soper et al. \cite{SoFi93} that speculations that the normal water
structure is significantly enhanced by the hydrophobic alkyl group is
groundless.  The calculations showed that methanol OH group is
strongly involved in hydrogen bonding, both as acceptor and as donor.
Analysis of the dynamics learns that the average residence time of a
water molecule in the first solvation shell is of the order of a few
picoseconds.  The vibrational spectrum of the water molecules is
hardly changed by the presence of the methanol, indicating that the
short-time dynamics is hardly affected by the presence of the methanol
molecule. Vibrational analysis shows that methanol OH-stretch peak is
a broad feature that is significantly red-shifted upon solvation,
confirming its hydrogen-bonding character. 

In conclusion, from comparison with available experimental data we
have shown that first-principle DFT-based molecular dynamics
simulation provides a reasonable accurate description of the structure
and dynamics of a dilute aqueous methanol solution. This opens the way
towards the study of chemistry involving methanol and larger alcohols in
water.
\vspace*{2mm}

\centerline{\bf Acknowledgements}
The Netherlands Organization for
Scientific Research is acknowledged financial support.  E.~J.~M.
acknowledges the "Royal Netherlands Academy of Arts and Sciences" for
financial support.
%\end{ack}

%\end{multicols}


\begin{thebibliography}{10}
\bibitem{CompResource}
Computational resources consisted of an IBM-SP and a cluster of
  state-of-the-art PC's. Calculations were executed in parallel using MPI and
  amounted to a total of $\approx$ 10000 hours of CPU-time.
\bibitem{ADF_basis}
Kohn-Sham orbitals are expanded in an uncontracted triple-$\zeta$ Slater type
  basis set augmented with 2p and 3d polarization functions for H and 3d and 4f
  polarization functions for C and O. The core states are kept frozen.
\end{thebibliography}
\end{document}